\documentclass[prc,aps,amsmath,amssymb,preprintnumbers,showpacs,twocolumn]{revtex4}
\usepackage{graphicx}% Include figure files
\usepackage{dcolumn}% Align table columns on decimal point
\usepackage{bm}% bold math

\def\thalf{{\textstyle{\frac{1}{2}}}}
\def\threehalf{{\textstyle{\frac{3}{2}}}}

\def\twothird{{\textstyle{\frac{2}{3}}}}

\newcommand{\bq}{\begin{eqnarray}}
\newcommand{\eq}{\end{eqnarray}}

\begin{document}
\preprint{NUC-MINN-04/6-T}
\title{Model Calculation of Effective Three-Body Forces}
\author{P.~J.~Ellis$^1$, T.~Engeland$^{2,3}$, M.~Hjorth-Jensen$^{2,3,4,5}$, 
M.~P.~Kartamyshev$^{2,3}$, and E.~Osnes$^{2,3}$} 
\affiliation{$^1$School of Physics and Astronomy, University of Minnesota, 
Minneapolis, MN 55455, USA\\$^2$Department of Physics, University of Oslo, 
N-0316 Oslo, Norway\\$^3$Center for Mathematics for Applications, 
University of Oslo, N-0316 Oslo, Norway\\
$^4$PH Division, CERN, CH-1211 Geneva 23, Switzerland\\
$^5$Department of Physics and Astronomy, Michigan State University,
East Lansing, MI 48824, USA}

\date{\today}

\begin{abstract}
We propose a scheme for extracting an effective three-body interaction 
originating from a two-nucleon interaction. This is based on the 
$\hat{Q}$-box method of Kuo and collaborators, where folded diagrams 
are obtained by differentiating a sum of non-folded diagrams with respect 
to the starting energy. To gain insight we have studied several examples
using the Lipkin model where the perturbative approach can be compared with
exact results. Numerically the three-body interactions can be significant 
and in a matrix example good accuracy was not obtained simultaneously for 
both eigenvalues with two-body interactions alone.
\end{abstract}
\pacs{21.60.Cs, 21.30.Fe, 24.10.Cn, 21.60.Fw}
\maketitle

%%%%%%%%%%%%%%%%%%%%%%

\section{Introduction}

The role of effective three-body forces in shell-model and nuclear structure 
calculations of medium-heavy and heavy nuclei is still an unsettled question. 
The question has received scattered attention over the years, but there has 
been no decisive study yet yielding a unique answer, see for example 
Refs.~\cite{herbert1,herbert2,herbert3,dehmo04}. Most of the studies 
seem to indicate that effective three-body forces play a rather modest role 
for nuclear spectra, although there is a possibility that binding energies 
may be more substantially affected. As 
pointed out by Zuker \cite{zuker1}, one is not able to obtain 
simultaneously a good reproduction of both the excitation spectra and 
the binding energy with effective interactions
based on two-body nucleon-nucleon interactions only, 
unless one fits an effective interaction to reproduce selected data, 
see for example the recent work of Otsuka {\em et al.} in 
Refs.~\cite{taka1,taka2}. To give an example, effective
interactions derived from two-body nucleon-nucleon interactions which fit 
nucleon-nucleon scattering data, are not able to reproduce the 
well-known shell closure   %pje spelling
in $^{48}$Ca nor     %pje
the excitation spectra of $^{47}$Ca and 
$^{49}$Ca \cite{zuker1,alex,oslo}. For the chain of oxygen 
isotopes one may even reach the conclusion that $^{28}$O is a bound 
nuclear system, unless  
the interaction is fitted to reproduce selected properties such
as binding energies and spectra 
of known and stable nuclei in the $1s0d$ shell \cite{alex}.  

On the other hand, theoretical interactions which have not been fitted 
do rather well in reproducing nuclear properties other than %pje
binding energies and shell closures.
Examples are nuclear spectra around $A\sim 100$ and $A\sim 132$
\cite{matej1,matej2,matej3,torgeir,jo,anne}, 
which come out surprisingly well in view of the 
complexity of the many-body problem. 
It has therefore been speculated that some of the abovementioned  deficiencies 
could be ascribed to the lack of three-body interactions, as seen in recent
Monte Carlo and no-core shell-model calculations of light nuclei with $A\le 16$
\cite{bob1,bob2,bob3,petr_erich2002,petr_erich2003}. 

In light nuclei the role of 
three-body forces seems rather well 
established. To reproduce the binding energy of the triton, for example, 
it is necessary to include a three-nucleon interaction in addition to 
the two-nucleon interaction. The situation in medium-heavy 
and heavy nuclei is however unclear, although similar arguments,
based on an analysis of $0p$, $1s0d$ and $1p0f$ nuclei by Zuker \cite{zuker1} 
lend support to the need for three-body interactions in  %pje small rewording
heavier nuclei as well.

In contrast to the case for light nuclei,
none of the present  calculations of effective interactions for medium and heavy nuclei
are 
sufficiently complete, however, to enable one to draw definite conclusions.
For nuclei with $A> 16$, one relies on shell-model analyses combined with appropriate effective interactions for a selected model space. 
Such interactions are typically based on perturbative many-body methods and  
because high-order calculations of the effective three-body contribution are 
prohibitively complicated, one is forced to limit oneself to low-order 
calculations. 
There are however classes of diagrams which can be summed to infinite order.
One of these classes is the set of folded diagrams.
The latter arise
due to the removal of the dependence on     %pje
the exact model-space energy
in the Brillouin-Wigner perturbation expansion. Through the $\hat{Q}$-box
formulation and its derivatives \cite{oslo,ko}, 
this set of diagrams can easily be summed
up.
With few exceptions \cite{herbert2}, three-body   %pje
folded diagrams have not been included. 
There are indications, however, that folded three-body diagrams may be 
important, even if non-folded ones are small. 

There is obviously a need to obtain a clearer 
picture of the role of effective three-body forces in heavier nuclei. Part 
of our motivation for studying the contribution of effective three-body 
forces comes from large-scale shell-model calculation of the entire range 
of Sn isotopes from A = 100 to A = 132, where the spectra are well 
reproduced whereas the binding energies are strongly overestimated. In 
fact, it was found that a small repulsive monopole contribution to the 
effective two-body interaction could cure this problem. The origin of
such a term is unknown, but it is speculated that it might be simulating 
an effective three-body interaction. Indeed Talmi \cite{igal} 
has shown that the binding energies of the entire range of Sn isotopes 
could be well fitted by an effective two-body plus an effective three-body 
interaction. In spite of the fact that both the Green's function Monte Carlo 
and no-core   %pje
shell-model approaches offer benchmark calculations of light nuclei 
with $A\le 16$ 
\cite{bob1,bob2,bob3,petr_erich2002,petr_erich2003}, and that coupled cluster
approaches can presently extend the region of ab initio  
calculations to $A\sim 40-56$ \cite{cc1}, 
features of heavier nuclei will most likely be studied 
within the framework of the shell model and appropriately defined 
effective interactions. Thus    %pje
we need to have a controlled approach which allows us to discern 
contributions from various many-body terms, such as effective three-body 
forces. 
Effective three-body forces may originate from three-nucleon forces as 
well as from two-nucleon forces. In order to be able to draw definite 
conclusions, we believe that it is necessary to study one effect at a time. 
Thus, before introducing a  three-nucleon interaction, we would study 
the contribution from effective three-body forces originating from the 
two-nucleon interaction. This is the topic of the present work. 

The paper is organized as follows. In Sec. \ref{secii} we propose an 
algorithm for evaluating the effective three-body interaction in the 
convenient $\hat{Q}$-box formalism introduced by Kuo and 
collaborators \cite{ko,oslo}. We prepare ourselves for this procedure by 
briefly summarizing the method for obtaining the effective two-body 
interaction. Because the method can be rather involved in 
realistic cases, we apply it in Sec. \ref{seciii} to a simple model,
namely the Lipkin model \cite{LMG65}, for which exact results can easily 
be obtained, see for example Ref.~\cite{vary04}. 
This is instructive both as regards the formalism and the 
numerical results.
In the final Sec. \ref{seciv} we draw 
conclusions and point to the need for more realistic calculations. 

%%%%%%%%%%%%%%%%%%%%%%

\section{The Effective One-, Two- and Three-Body Interactions\label{secii}}

The Hamiltonian acting in the complete Hilbert space (usually 
infinite) consists of an unperturbed one-body part, $H_0$, and a 
perturbation ${\cal V}$, namely $H=H_0+{\cal V}$. The goal is to obtain
an effective interaction, $V$, such that $H=H_0+V$ acting in a chosen 
model or valence
space yields a set of eigenvalues which are identical to (a subset of)
those of the complete problem. To this end  Kuo and collaborators
introduced the $\hat{Q}$-box, see Refs. \cite{oslo,ko}, which, in 
principle, contains all non-folded diagrams which are attached to
the valence particle lines. Unlinked pieces which refer to excitations 
of a closed-shell core are absent since the energy is calculated relative 
to the closed-shell ground state \cite{bran}. The effective interaction also 
requires that folded diagrams be included to all orders and this can 
be formally written
\begin{equation}
V=\lim_{n\rightarrow\infty}V(n)\;, \label{veff}  %pje moved label
\end{equation}
where
\begin{equation}
V(n)=\hat{Q}(\omega)+\sum_{m=1}^\infty\frac{1}{m!}\frac{d^m\hat{Q}}
{d\omega^m}\bigl\{V(n-1)\bigr\}^m\;, \label{veff2}  %pje new label
\end{equation}
with $V(0)\equiv\hat{Q}$ and matrix multiplication implied.
Here $\omega$ is the unperturbed energy of the initial state. In order
to perform calculations with $N$ particles in the model space it is
necessary to separate the effective interaction $V$ into its one-, two-
and three-body components (in principle up to $N$-body interactions are
needed, but we shall not go beyond three-body). This 
brief sketch establishes our notation; a complete discussion is to be 
found in Refs. \cite{oslo,ko,bran,usrev}. 

\begin{figure}[htbp]
 \includegraphics[width=8truecm]{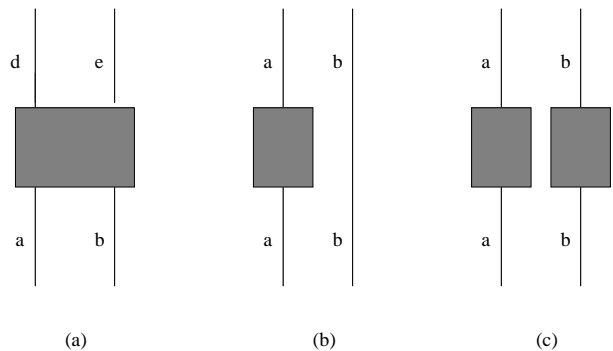}
\caption{Two-body diagrams discussed in the text.} \label{f1}
\end{figure}

Consider first the case where our model space consists of a single particle,
$N=1$;
we assume that a given total angular momentum $j$ is represented only once.
Then Eq. (\ref{veff}) can be solved for each state separately, yielding
the complete folded series for the one-body effective interaction
$V\equiv V^{(1)}$. The one-body effective Hamiltonian is thus 
$H_0+V^{(1)}$.

Now consider the case where we have two particles in the model space, $N=2$.
The $\hat{Q}$-box again consists  of all non-folded diagrams. It will 
include diagrams such that the interactions link the two valence 
lines together as indicated schematically in Fig. \ref{f1}(a), diagrams where
one line is non-interacting as in Fig. \ref{f1}(b) and also diagrams where
the interactions on each line are separate so that the diagram is 
valence unlinked as in Fig. \ref{f1}(c). All of these components need to 
be included so as to generate the cross terms in the full folded effective
interaction of Eqs. (\ref{veff}) and (\ref{veff2}). %pje
Valence unlinked diagrams are removed by
the folding process so that
the effective interaction is completely linked \cite{bran}.
Now because of the one- plus two-body nature of the $\hat{Q}$-box, the 
result for $V$ from Eq. (\ref{veff}) will contain linked, folded 
two-body components which are schematically of the form of Fig. \ref{f1}(a) 
plus linked, folded 
one-body components where one line is non-interacting, as in 
Fig. \ref{f1}(b). The one-body piece $V^{(1)}$ is already determined, 
as indicated above, so the purely two-body interaction $V^{(2)}$ can be 
obtained from
\begin{equation}
\langle de|V|ab\rangle_2=\langle de|V^{(2)}|ab\rangle 
+\delta_{da}\delta_{eb}\Bigl(\!\langle a|V^{(1)}|a\rangle
+\langle b|V^{(1)}|b\rangle\!\Bigr). \label{whole2}
\end{equation}
The subscript on the left indicates the value of $N$ for which 
Eq. (\ref{veff}) is solved.
We assume here that the initial and final states in Fig. \ref{f1}(a) are
in a standard order so that $e=a$ implies $d\neq b$. Here the two-body 
matrix elements are ``antisymmetrized'', i.e $\langle de|V|ab\rangle$
implies $\langle de|V|ab-ba\rangle$. They are not coupled to a total
angular momentum or isospin since our shell model calculations are 
carried out in the $m$-scheme.

\begin{figure}[t]    
 \includegraphics[width=7.5truecm]{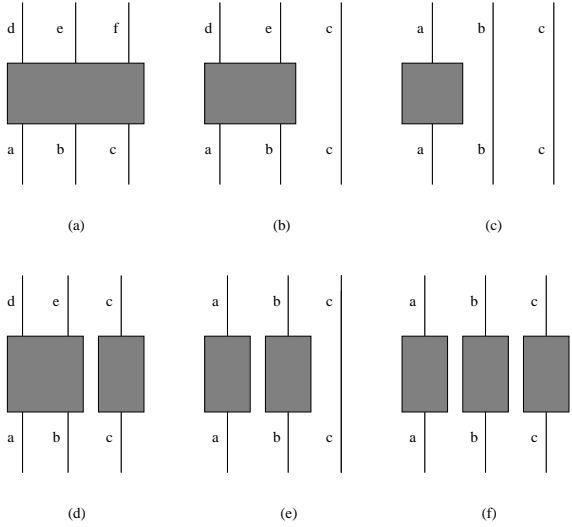}
\caption{Three-body diagrams discussed in the text.} \label{f2}
\end{figure}

Turning to the more interesting three-body case ($N=3$), consider the 
$\hat{Q}$-box. It contains diagrams such that the interactions link 
the three valence lines together as indicated schematically in 
Fig. \ref{f2}(a), diagrams where two of the lines are linked 
(Fig. \ref{f2}(b)) or where only one line is interacting as in 
Fig. \ref{f2}(c). The valence unlinked diagrams can involve a two-body
and a one-body piece (Fig. \ref{f2}(d)) or two or three one-body pieces
(Figs. \ref{f2}(e) and (f)). Notice that there are no valence unlinked 
diagrams if, as is often the case, one-body interactions are excluded.
In general all of the components illustrated in Fig. \ref{f2} need to 
be included in generating the full folded, valence-linked effective 
interaction of Eqs. (\ref{veff}) and (\ref{veff2}). %pje
The interaction has a three-body 
contribution where the three lines are linked together, schematically in 
the form of Fig. \ref{f2}(a), two-body interactions acting on all possible
pairs of lines (Fig. \ref{f2}(b)) and one-body interactions 
acting each line separately (Fig. \ref{f2}(c)). Using Eq. (\ref{whole2})
we can disentangle the desired $V^{(3)}$ from 
\begin{widetext}
\bq
\langle def|V|abc\rangle_3&=&\langle def|V^{(3)}|abc\rangle
+\delta_{fc}\langle de|V|ab\rangle_2-\delta_{fb}
\langle de|V|ac\rangle_2+\delta_{fa}\langle de|V|bc\rangle_2
+\delta_{eb}\langle df|V|ac\rangle_2\nonumber\\
&&-\delta_{ec}\langle df|V|ab\rangle_2-\delta_{ea}
\langle df|V|bc\rangle_2+\delta_{da}\langle ef|V|bc\rangle_2
-\delta_{db}\langle ef|V|ac\rangle_2
+\delta_{dc}\langle ef|V|ab\rangle_2\nonumber\\
&&
-\delta_{da}\delta_{eb}\delta_{fc}
\Bigl(\langle a|V^{(1)}|a\rangle+\langle b|V^{(1)}|b\rangle
+\langle c|V^{(1)}|c\rangle\Bigr)\;. \label{whole3}
\eq
\end{widetext}
Again the subscript on the left indicates the value of $N$ for which 
Eq. (\ref{veff}) is solved. Of course for a given matrix element only 
a few of the delta functions will be nonzero.
Since the inital and final states are in a chosen standard order
the one-body term only contributes when $d=a$, $e=b$ and $f=c$.
Here the uncoupled three-body matrix elements are ``antisymmetrized'', 
i.e $\langle def|V|abc\rangle$
implies 
\begin{displaymath}
\langle def|V|abc-acb+cab-bac+bca-cba\rangle\;,
\end{displaymath}
labelling in order for particle numbers 1, 2 and 3. Equation 
(\ref{whole3}) has one-body components in $\langle V\rangle_2$
as well as a compensating explicit term. By using 
Eq. (\ref{whole2}) the equation can be cast just in terms of one-, 
two-, and three-body effective interactions:
\begin{widetext}
\bq
\langle def|V|abc\rangle_3&=&\langle def|V^{(3)}|abc\rangle
+\delta_{fc}\langle de|V^{(2)}|ab\rangle-\delta_{fb}
\langle de|V^{(2)}|ac\rangle+\delta_{fa}\langle de|V^{(2)}|bc\rangle
+\delta_{eb}\langle df|V^{(2)}|ac\rangle\nonumber\\
&&-\delta_{ec}\langle df|V^{(2)}|ab\rangle-\delta_{ea}
\langle df|V^{(2)}|bc\rangle+\delta_{da}\langle ef|V^{(2)}|bc\rangle
-\delta_{db}\langle ef|V^{(2)}|ac\rangle
+\delta_{dc}\langle ef|V^{(2)}|ab\rangle\nonumber\\
&&
+\delta_{da}\delta_{eb}\delta_{fc}
\Bigl(\langle a|V^{(1)}|a\rangle+\langle b|V^{(1)}|b\rangle
+\langle c|V^{(1)}|c\rangle\Bigr)\;. \label{whole3p}
\eq
\end{widetext}
This just spells out explicitly the fact that the complete effective
interaction requires all possible valence-linked contributions.

Thus, given the $\hat{Q}$-box in the one, two and three particle systems
we can obtain the one-, two- and three-body effective interactions. 
Obviously it would be possible to generalize this procedure to obtain 
four-body and higher interactions.   %pje
However, since almost all calculations 
have stopped at the two-body level, it is sensible to investigate just 
three-body interactions. Here we attempt to get a feel for their impact 
in a simple model situation.

%%%%%%%%%%%%%%%%%%%%%%
\section{Application to the two-level Lipkin model\label{seciii}}

The Lipkin model \cite{LMG65} consists of two single-particle levels
labelled by $\sigma=-$ and $+$, each of which has a degeneracy $p$. We write
the Hamiltonian
\begin{widetext}
\vspace{-.7cm}
\bq
H&=& H_0+{\cal V}\;,\quad{\rm where}\nonumber\\
H_0&=&\thalf\xi\sum_{p\sigma}{\sigma}a_{p\sigma}^{\dagger}a_{p\sigma}\;,
\quad{\rm and}\nonumber\\
{\cal V}&=&\thalf V\sum_{pp'\sigma}a^{\dagger}_{p\sigma}
a^{\dagger}_{p'\sigma}a_{p'-\sigma}a_{p-\sigma}+
\thalf W\sum_{pp'\sigma}a^{\dagger}_{p\sigma}a_{p'-\sigma}^{\dagger}
a_{p'\sigma}a_{p-\sigma}
+\thalf U\sum_{pp'\sigma}\left[a^{\dagger}_{p\sigma}
a_{p'\sigma}^{\dagger}
a_{p'-\sigma}a_{p\sigma}+a^{\dagger}_{p\sigma}a_{p'-\sigma}^{\dagger}
a_{p'\sigma}a_{p\sigma}\right].
\eq
\end{widetext}
\vspace{-.2cm}
Here $H_0$ is the unperturbed  Hamiltonian with single particle energies
$\pm\thalf\xi$. The two-body interaction, ${\cal V}$, has three terms. The
interaction $V$ acts between a pair of particles with parallel spins and
changes the spins from $++$ to $--$, or vice versa. The interaction $W$
is a spin-exchange interaction and $U$, which was not present in the original
model \cite{LMG65}, flips the spin of one particle.
It is of interest to note that the interaction does not change the value of
the degeneracy labels $pp'$.

Since each particle has only two possible states, the use of the quasi-spin
formulation was suggested by Lipkin {\it et al.} \cite{LMG65}. The
quasi-spin operators obey angular momentum commutation relations and are
defined by
\begin{equation}
J_z=\thalf\sum_{p\sigma}\!{\sigma}a^{\dagger}_{p\sigma}a_{p\sigma}\,,\,
J_+=\sum_p\!a_{p+}^{\dagger}a_{p-}\,,\,J_-=\sum_p\!a_{p-}^{\dagger}a_{p+}\,.
\end{equation}
The Hamiltonian can then be compactly expressed in the form
\begin{eqnarray}
H&=&{\xi}J_z+\thalf V(J_+^2+J_-^2)+\thalf W(J_+J_-+J_-J_+-n)\nonumber\\
&&+\thalf U(J_++J_-)(n-1)\;,
\label{LIPH}
\end{eqnarray}
where the number operator $n=\sum_{p\sigma}a^{\dagger}_{p\sigma}
a_{p\sigma}$.
The operator $J^2=\frac{1}{2}(J_+J_-+J_-J_+)+J_z^2$ commutes with the
Hamiltonian so the Hamiltonian matrix breaks up into
submatrices of dimension $2J+1$, each associated with different values of
$J$; for a given number of particles $N$ the largest angular momentum
corresponds to $J=\thalf N$. 

Here we need the case $N=3$ and for our purposes it is sufficient
to consider $J=\threehalf$.
We denote the basis state for three particles by
\begin{equation}
|p_1\pm,p_2\pm,p_3\pm\rangle=a^\dagger_{p_1\pm}a^\dagger_{p_2\pm}
a^\dagger_{p_3\pm}|0\rangle\;,
\end{equation}
where $p_i$ refers to the degeneracy label and $|0\rangle$ is the 
vacuum state. Then the basis states for the $J=\threehalf$ matrix are:
\begin{widetext}
\vspace{-.7cm}
\bq
|1\rangle\equiv|J=\threehalf J_z=-\threehalf\rangle
&=&|p_1-,p_2-,p_3-\rangle\;,\nonumber\\
|2\rangle\equiv|J=\threehalf J_z=-\thalf\rangle
&=&\bigl\{|p_1-,p_2-,p_3+\rangle
+|p_1-,p_2+,p_3-\rangle+|p_1+,p_2-,p_3-\rangle\bigr\}/\sqrt{3}
\;,\nonumber\\
|3\rangle\equiv|J=\threehalf J_z=+\thalf\rangle
&=&\bigl\{|p_1-,p_2+,p_3+\rangle
+|p_1+,p_2-,p_3+\rangle+|p_1+,p_2+,p_3-\rangle\bigr\}/\sqrt{3}
\;,\nonumber\\
|4\rangle\equiv|J=\threehalf J_z=+\threehalf\rangle
&=&|p_1+,p_2+,p_3+\rangle\;.\label{3basis}
\eq
\end{widetext}
\vspace{-.2cm}
In this basis the Hamiltonian matrix is 
\begin{equation}
\left(\begin{array}{cccc}
-\threehalf\xi&\sqrt{3}U&\sqrt{3}V&0\\[0.5mm]
\sqrt{3}U&-\thalf\xi+2W&2U&\sqrt{3}V\\
\sqrt{3}V&2U&\thalf\xi+2W&\sqrt{3}U\\
0&\sqrt{3}V&\sqrt{3}U&\threehalf\xi
\end{array}\right)\;. \label{mat44}
\end{equation}

\subsection{The Case $U=0$\label{subsecu0}}

We consider first the simplest case where $U=0$ so that the matrix 
(\ref{mat44}) splits into two $2\times2$ matrices. Consider the matrix
formed by states $|1\rangle$ and $|3\rangle$:
\begin{equation}
\left(\begin{array}{cc}
-\threehalf\xi&\sqrt{3}V\\[0.5mm]
\sqrt{3}V&\thalf\xi+2W
\end{array}\right)\;, \label{mat22}
\end{equation}
whose eigenvalues are
\begin{equation}
\lambda_\pm^{(3)}=-\threehalf\xi
+(\xi+W)\left[1\pm\sqrt{1+\frac{3V^2}{(\xi+W)^2}}\;\right]\;.
\label{ex3u0} 
\end{equation}
In principle both eigenvalues are given by an exact treatment 
of the perturbation series \cite{suz} since Eq. (\ref{veff}) is 
simply a rearrangement of the 
exact eigenvalue equation. In practice approximations imply that only
one eigenvalue is obtained and we will focus on $\lambda_-^{(3)}$ using
the model space state $|1\rangle$. 

\begin{figure}[t]
 \includegraphics[width=8truecm]{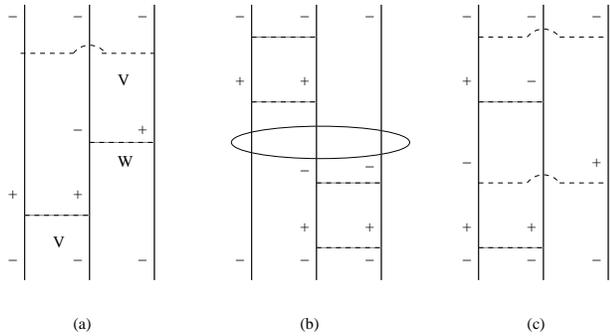}
\caption{Third-order (a) and fourth-order ((b) and (c)) three-body 
diagrams discussed in the text. Diagram (b) is a folded diagram drawn
in unfolded form for clarity.} \label{f3}
\end{figure}

Now directly from the matrix (\ref{mat22}), or by explicitly drawing 
diagrams, the second order $\hat{Q}$-box is $3V^2/(-2\xi)$. Clearly 
the $W$ interaction can be incorporated to all orders yielding an exact 
$\hat{Q}$-box of $3V^2/(-2\xi-2W)$. The folded diagrams can then be
included by differentiation as indicated in Eq. (\ref{veff2}). In the 
approximation that we stop at order $V^6$  the interaction is
\begin{eqnarray}
\langle V\rangle_3&=&-\frac{3V^2}{2(\xi+W)}
+\frac{9V^4}{8(\xi+W)^3}-\frac{27V^6}{16(\xi+W)^5}\nonumber\\
&\equiv& f_3(\xi)\;.
\end{eqnarray}
In order to disentangle the three-body interaction we need the two-body 
interaction for the state $|p_1-,p_2-\rangle$. Since
this is only coupled to $|p_1+,p_2+\rangle$ by the Hamiltonian, the 
necessary matrix is
$\left(\begin{array}{cc}
-\xi&V\\
V&\xi
\end{array}\right)$.  For later use we note that the exact lowest 
eigenvalue is $\lambda_-^{(2)}=-\xi\sqrt{1+\frac{V^2}{\xi^2}}$.
Now the exact $\hat{Q}$-box is $V^2/(-2\xi)$ and carrying out the
folding to order $V^6$ as before 
\begin{equation}
\langle V\rangle_2=-\frac{V^2}{2\xi}+\frac{V^4}{8\xi^3}
-\frac{V^6}{16\xi^5}
\equiv f_2(\xi)\;. \label{u02bord}
\end{equation}
The one particle wavefunction is $a^\dagger_{p-}|0\rangle$ with energy 
$-\thalf\xi$ given by the unperturbed Hamiltonian $H_0$, so that the
one-body interaction $V^{(1)}$ is zero. Thus, using Eq. (\ref{whole3}) 
or (\ref{whole3p}), and noting that because of the degeneracy label $p$ 
there will only be three contributions from the two-body terms
\begin{widetext}
\begin{equation}
\langle V^{(3)}\rangle=\langle V\rangle_3-3\langle V\rangle_2
=\frac{3V^2W}{2\xi^2}+\frac{3V^2}{4\xi^3}(V^2-2W^2)  
+\frac{3V^2W}{8\xi^4}(4W^2-9V^2)-\frac{3V^2}{4\xi^5}
(2V^4-9V^2W^2+2W^4)\;. \label{u04ord}
\end{equation}
\end{widetext}
Here we have expanded to sixth order in the interaction. Notice that there
is no second-order contribution. This result can also be obtained by
evaluating diagrams, for brevity we do not go beyond fourth order.
In third order the diagram of Fig. \ref{f3}(a) is required, while in
fourth order the folded diagram of  Fig. \ref{f3}(b), drawn in unfolded form, 
and the non-folded diagram (c) are needed; by counting the
number of independent diagrams of each type the numerical factors in 
Eq. (\ref{u04ord}) are obtained. If $W$ is set to zero in %pje sentences added
Eq. (\ref{u04ord}) the remaining three-body contributions come from  folded
diagrams, Fig. \ref{f3}(b) and higher orders. The coefficients of the 
order $V^4$ and $V^6$ terms in Eq. (\ref{u04ord}) are larger than three 
times the corresponding values in  Eq. (\ref{u02bord}). This supports the 
suggestion that even if non-folded three-body diagrams are small, three-body 
contributions arising from folding could be important.

We can, of course, obtain the exact three-body interaction for this 
simple Hamiltonian using the exact eigenvalues given. As we have remarked,
the one-body interaction $V^{(1)}$ is zero. Then, subtracting the 
unperturbed energy, the two-body effective interaction is
\begin{equation}
\langle V^{(2)}\rangle=\xi\left(1-\sqrt{1+\frac{V^2}{\xi^2}}\right)\;.
\label{2bu0}
\end{equation}
The desired exact three-body interaction is therefore
\begin{equation}
\langle V^{(3)}\rangle=\lambda_-^{(3)}-3\langle V^{(2)}\rangle
+\threehalf\xi\;,\label{ex3}
\end{equation}
where the last term removes the contribution from $H_0$
in $\lambda_-^{(3)}$. Expanding this expression to sixth order of course 
yields Eq. (\ref{u04ord}) again.
Notice that since $V^{(3)}$ depends on $W$ while $V^{(2)}$ does not,
suitable adjustment of this parameter could yield a three-body 
interaction of magnitude comparable to the two-body interaction. For
instance, in the limit $W\rightarrow\infty$ we have 
$V^{(3)}\rightarrow-3V^{(2)}$
since the eigenvalue is simply $-\threehalf\xi$ in this case.

It is instructive to make the model a little more elaborate by changing 
the parameter $\xi$ of the one-body Hamiltonian to $\xi+\epsilon$. The 
first term is included in $H_0$ as before and $\epsilon$ is treated
as a perturbation. The matrix (\ref{mat22}) then becomes
\begin{equation}
\left(\begin{array}{cc}
-\threehalf\xi-\threehalf\epsilon&\sqrt{3}V\\
\sqrt{3}V&\thalf\xi+\thalf\epsilon+2W
\end{array}\right)\;. \label{mat22p}
\end{equation}
\begin{figure}[t]
 \includegraphics[width=8truecm]{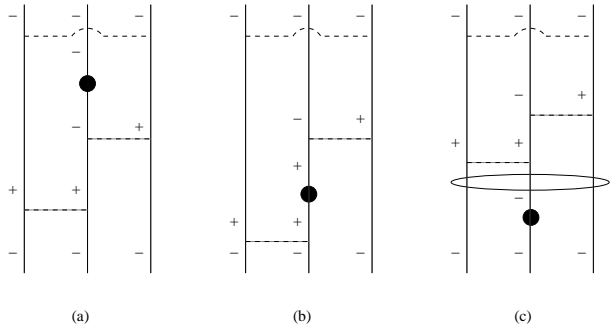}
\caption{Fourth-order three-body diagrams discussed in the text. 
Diagram (c) is a folded diagram drawn in unfolded form for clarity
and the dot represents a one-body insertion.} \label{f4}
\end{figure}
 From the matrix directly, or by drawing diagrams, the 
$\hat{Q}$-box through second order contains both one- and two-body parts,
namely $-\threehalf\epsilon-3V^2/(2\xi)$. Starting in third order
there will be diagrams with $\epsilon$ insertions. Since the intermediate
state always involves two $+$ particles and one $-$ particle, the net 
insertion is $+\thalf\epsilon$ (note that valence unlinked contributions
of the form of Fig. \ref{f2}(d) occur here). If these insertions, together
with the $W$ interactions are summed to all orders one obtains
$-\threehalf\epsilon-3V^2/(2\xi+2W+\thalf\epsilon)$, a result that can be
obtained immediately from the matrix (\ref{mat22p}). Now an $\epsilon$ 
insertion on each of the three $-$ particles of the model space state 
$|1\rangle$ gives a net contribution of $-\threehalf\epsilon$. If these 
insertions are folded out to all orders one obtains
$-\threehalf\epsilon-3V^2/(2\xi+2W+2\epsilon)$, in the process removing
valence-unlinked diagrams. This result again follows immediately from the 
matrix (\ref{mat22p}). If this modified $\hat{Q}$-box is used in 
Eq. (\ref{veff2}) to generate all the additional folded diagrams required
we obtain 
$\langle V\rangle_3=-\threehalf\epsilon+f_3(\xi+\epsilon)$. A similar 
procedure in the two-body case gives 
$\langle V\rangle_2=-\epsilon+f_2(\xi+\epsilon)$.
In the one-body case $\langle V^{(1)}\rangle=-\thalf\epsilon$.
Then using Eq. (\ref{whole3}) 
\begin{widetext}
\bq
\langle V^{(3)}\rangle&=&\langle V\rangle_3-3\langle V\rangle_2
+3\langle V^{(1)}\rangle\nonumber\\
&=&\frac{3V^2W}{2\xi^2}+\frac{3V^2}{4\xi^3}(V^2-2W^2-4\epsilon W)
+\frac{3V^2}{8\xi^4}(4W^3-9V^2W-6V^2\epsilon+12W^2\epsilon+12W\epsilon^2)
\nonumber\\
&&-\frac{3V^2}{4\xi^5}(2V^4-9V^2W^2+2W^4-18V^2W\epsilon+8W^3\epsilon
-6V^2\epsilon^2+12W^2\epsilon^2+8W\epsilon^3)\;, \label{u04ordp}
\eq
\end{widetext}
to sixth order in the interaction. Obviously the terms which do not
involve $\epsilon$ are the same as in Eq. (\ref{u04ord}). Notice
that terms of the form $V^2\epsilon^n/\xi^{n+1}$ necessarily belong
to the two-body interaction and do not occur here.
The fourth order term involving  $\epsilon$ is generated from the 
diagrams of Fig. \ref{f4}(a) and (b), as well as the folded diagram (c). 
Here $V^{(1)}$ is represented by a dot. Each diagram stands for 
several different diagrams of the same general structure and the coefficient
in Eq. (\ref{u04ordp}) is obtained by accounting for all of them.
The result in Eq. (\ref{u04ordp}) can also be obtained
by a power series expansion of the exact result which is the analogue of
Eq. (\ref{ex3}), namely
\begin{equation}
\langle V^{(3)}\rangle=\lambda_-^{(3)}-3\langle V^{(2)}\rangle
-3\langle V^{(1)}\rangle+\threehalf\xi\;,\label{ex4}
\end{equation}
with the replacement $\xi\rightarrow\xi+\epsilon$ being made in the first 
two terms on the right.

\subsection{The Case $U\neq0$\label{subsecu1}}

\begin{figure}[t]
 \includegraphics[width=7truecm]{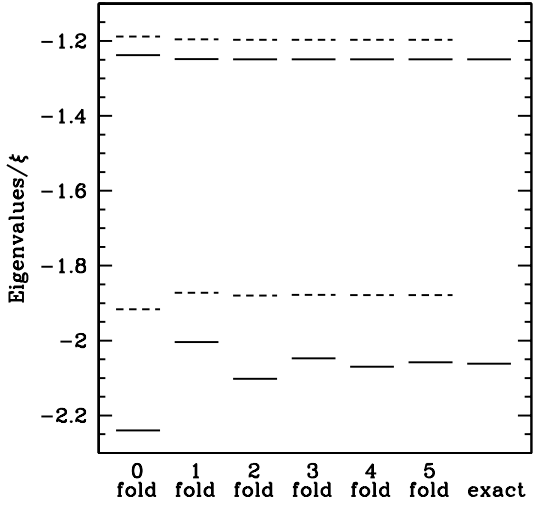}
\caption{Comparison of the eigenvalues (in units of $\xi$) of the 
effective interaction with $n$ folds to the exact result. For the  
solid lines two- and three-body interactions are included, while 
for the dashed lines only two-body interactions are considered.} 
\label{f5}
\end{figure}

Here we take the model space to consist of states $|1\rangle$ and
$|2\rangle$ of Eq. (\ref{3basis}) and require our effective interaction
to reproduce two of the eigenvalues of the matrix (\ref{mat44}). It is 
simplest to work with a degenerate model space and we choose the 
unperturbed energy to be $-\threehalf\xi$. The remainder of the 
diagonal $(2,2)$ contribution
in (\ref{mat44}) is then included as a perturbation. The lowest order 
$\hat{Q}$-box of a given type is easily written down as discussed in 
Sec. \ref{subsecu0}. Here we also need to account for multiple scatterings 
back and forth between states $|3\rangle$ and $|4\rangle$ which is 
easily done by summing the infinite series. Thus we can obtain the
exact $\hat{Q}$-box:
\bq
&&\langle1|\hat{Q}|1\rangle=\frac{3V^2}{-2\xi-2W+U^2/\xi}\;,\nonumber\\
&&\langle2|\hat{Q}|1\rangle=\langle1|\hat{Q}|2\rangle=\sqrt{3}U+
\frac{2\sqrt{3}UV}{-2\xi-2W+U^2/\xi}\nonumber\\
&&\qquad\qquad\qquad\qquad\ \ +\frac{\sqrt{3}UV^2}{2\xi(\xi+W)-U^2}
\;,\nonumber\\
&&\langle2|\hat{Q}|2\rangle=\xi+2W+\frac{4U^2}{-2\xi-2W+U^2/\xi}
\nonumber\\
&&\quad+\frac{V^2}{-\xi+U^2/(2\xi+2W)}+\frac{4VU^2}{2\xi(\xi+W)-U^2}\;.
\label{q4u}
\eq
Insertion of this in Eq. (\ref{veff2}) will yield an effective interaction
which reproduces two of the exact eigenvalues.
This is illustrated by the solid lines in Fig. \ref{f5}. We work in units 
of $\xi$ and have 
chosen the dimensionless ratios  $V/\xi=W/\xi=-0.4$ and $U/\xi=-0.15$. 
The label 0 fold in Fig. \ref{f5} implies that the solid lines are obtained 
from the $\hat{Q}$-box alone, while the labelling $n$ fold implies that
the summation over $m$ in Eq. (\ref{veff2}) extends to $m=n$. As $n$ is 
increased so that more folded diagrams are included the eigenvalues of the 
effective interaction converge to the exact result. The deviations 
are less than 0.2\% by the time $n=5$.

\begin{figure}[t]
 \includegraphics[width=7truecm]{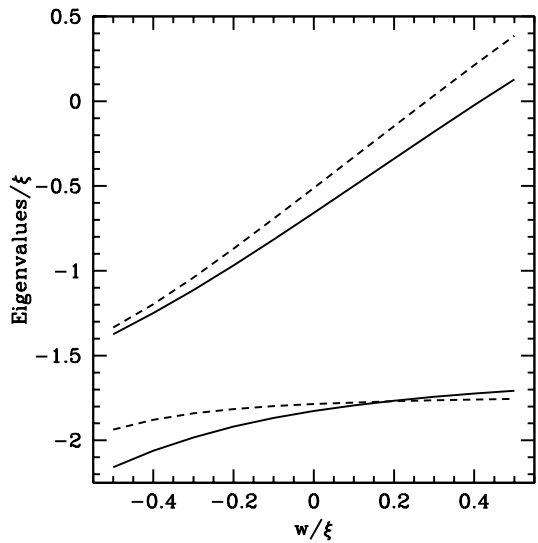}
\caption{Eigenvalues (in units of $\xi$) obtained with the two-body
effective interaction (dashed curves) compared with the exact values
(solid curves), as a function of $w/\xi$.} \label{f7}
\end{figure}

This analysis necessarily includes effective three-body interactions and we 
would like to assess the effect of their removal. To that end consider first 
the two-body problem. We only need to consider the $J=1$ matrix which is
\begin{equation}
\left(\begin{array}{ccc}
-\xi&U/\sqrt{2}&V\\
U/\sqrt{2}&W&U/\sqrt{2}\\
V&U/\sqrt{2}&\xi
\end{array}\right)\;, \label{mat33}
\end{equation}
in the basis
\bq
|a\rangle&\equiv&|J=1 J_z=-1\rangle=|p_1-,p_2-\rangle\;,\nonumber\\
|b\rangle&\equiv&|J=1 J_z=0\rangle=\bigl\{|p_1-,p_2+\rangle
+|p_1+,p_2-\rangle\bigr\}/\sqrt{2}\;,\nonumber\\
|c\rangle&\equiv&|J=1 J_z=+1\rangle
=|p_1+,p_2+\rangle\;.\label{2basis}
\eq
Taking the model space to comprise states $|a\rangle$ and $|b\rangle$
and choosing degenerate unperturbed energies of $-\xi$, the exact
$\hat{Q}$-box is
\begin{eqnarray}
\langle a|\hat{Q}|a\rangle&=&-\frac{V^2}{2\xi}\;,\nonumber\\
\langle b|\hat{Q}|a\rangle&=&\langle a|\hat{Q}|b\rangle=\frac{U}{\sqrt{2}}
-\frac{VU}{2\sqrt{2}\xi}\;,\nonumber\\
\langle b|\hat{Q}|b\rangle&=&\xi+W-\frac{U^2}{4\xi}\;.
\end{eqnarray}
\begin{figure}[t]
 \includegraphics[width=3truecm]{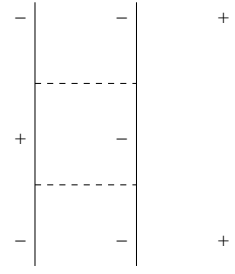}
\caption{A three-body diagram discussed in the text.} \label{f6}
\end{figure}
When used in Eq. (\ref{veff2}) this will give an effective interaction which
yields two of the exact two-body eigenvalues.
However our interest is in using this in the three 
particle case. The $(a,a)$ element will contribute thrice to the
$(1,1)$ component of the three body $\hat{Q}$-box and once to the
$(2,2)$ case. Note that for the latter the use of  $-\threehalf\xi$
for the unperturbed energies requires that the energy denominator
be modified to $3\xi$. This can be viewed as arising from diagrams
with a one-body insertions of $-\xi$ folded out to all
orders to correct the energy of the model space $+$ state. 
The two-body part of the  $(b,b)$ element
contributes twice to the $(2,2)$ component; the one-body part is 
associated with the single $+$ state and therefore only contributes 
once. Finally the $(b,a)$ element contributes thrice to the $(2,1)$ 
component multiplied by a factor $\sqrt{\twothird}$ due to the 
normalization in 
Eqs. (\ref{3basis}) and (\ref{2basis}). In this way the one- plus 
two-body contribution to the three-body $\hat{Q}$-box is found to be
\begin{eqnarray}
\langle1|\hat{Q}|1\rangle&=&-\frac{3V^2}{2\xi}\;,\nonumber\\
\langle2|\hat{Q}|1\rangle&=&\langle1|\hat{Q}|2\rangle=\sqrt{3}U-
\frac{\sqrt{3}VU}{2\xi} \;,\nonumber\\
\langle2|\hat{Q}|2\rangle&=&\xi+2W-\frac{U^2}{2\xi}
-\frac{V^2}{3\xi}\;. \label{q2u}
\end{eqnarray}
Note a subtlety here. A diagram of the type shown in Fig. \ref{f6} 
might appear to be a two-body contribution to the three-particle 
interaction. However if the non-interacting line is erased the 
remainder cannot contribute to the two-body interaction because the 
intermediate state is in the model space. This is no longer the case 
with three particles present where the intermediate state lies outside 
the model space, thus it is properly counted as a three-body 
interaction. Comparing Eq. (\ref{q2u}) to Eq. (\ref{q4u}) we see that 
here the three-body interactions begin at second order due to the 
presence of state $|2\rangle$. 
Using the two-body approximation of Eq. (\ref{q2u}) for the $\hat{Q}$-box
in Eq. (\ref{veff2}) yields the eigenvalues denoted by dashed lines 
in Fig. \ref{f5}. Comparison with the exact result shows that the two-body 
approximation converges more rapidly in terms of the number of folds, 
but the converged result has a lower (upper) eigenvalue which is in error
by 9\% (4\%); this is not insignificant. The point is also made 
in Fig. \ref{f7} where the same values of $V/\xi$ and $U/\xi$ are used,
but the parameter $W/\xi$ is allowed to vary. Again the solid line
employs an effective interaction with one-, two- and %pje slight rewording
three-body terms, while the dashed line is obtained when the three-body 
contributions are omitted. 
Clearly the accuracy of the two-body approximation is dependent on 
the precise values of the parameters used. In this example it can be 
quite accurate for one of the eigenvalues for particular values
of $W/\xi$, but then it is inaccurate for the other eigenvalue.

%

%%%%%%%%%%%%%%%%%%%%%%%%%%%
\section{Concluding remarks\label{seciv}}
 
We have shown how the effective three-body interaction may be isolated 
from the nonfolded and folded diagrams generated from a $\hat{Q}$-box 
which contains one-, two- and three-body terms. In order to gain some 
insight this was applied to the Lipkin model where exact results can easily 
be obtained. Even for this simple example subtle points arose.
Numerically we showed that for a system of three particles the
three-body components of the effective interaction cannot be assumed to be
negligible. Interestingly in our effective $2\times2$ matrix example
the two-body interaction alone sometimes gave quite an accurate 
result for one of the eigenvalues, but then the other was rather inaccurate.

This suggests that the role of three-body interactions in realistic 
situations deserves further study, in particular in medium-heavy  and heavy nuclei. 
As mentioned in the Introduction we hope 
to carry out such a study for the entire range of Sn isotopes using the 
formalism discussed here. The difficulty in explaining the trend of the
binding energies is a particular motivation, but more important is the fact 
that three-body effects scale with the number of valence particles cubed, 
while two-body contributions scale with the square. Thus an examination of 
some thirty Sn isotopes should allow a rather definitive assessment of 
the role of three-body interactions to be made.  
  
%%%%%%%%%%%%%%%%%%%%%%%%%%%
\begin{acknowledgments}
This work was supported in part by the US Department of Energy under 
grant DE-FG02-87ER40328 and the Research Council of Norway.
\end{acknowledgments}

%%%%%%%%%%%%%%%%%%%%%%%%%%%


\begin{thebibliography}{99}
\bibitem{herbert1} 
A.\ Polls, H.\ M\"{u}ther, A.\ Faessler,
T.~T.~S.\ Kuo, and E.\ Osnes, Nucl.\ Phys.\
 {\bf A401}, 124  (1983).
\bibitem{herbert2} H.\ M\"{u}ther, A.\ Polls, and T.~T.~S.\ Kuo, 
Nucl.~Phys.~{\bf A435}, 548 (1985).
\bibitem{herbert3} P.~K.~Rath, A.~Faessler, H.~M\"uther, and A.~Watt,
Nucl.~Phys.~{\bf A492}, 127 (1989).
\bibitem{dehmo04} D.~J.~Dean, T.~Engeland, M.~Hjorth-Jensen, M.~P.~Kartamyshev,
and E.~Osnes, Prog.~Part.~Nucl.~Phys.~{\bf 53}, 419 (2004).
\bibitem{zuker1} A.~P.~Zuker, Phys.~Rev.~Lett.~{\bf 90}, 042502 (2003).
\bibitem{taka1} M.~Homna, T.~Otsuka, B.~A.~Brown, and T.~Mizusaki, Phys.~Rev.~C
{\bf 69}, 034335 (2004).
\bibitem{taka2} T.~Otsuka, M.~Homna, T.~Mizusaki, N.~Shimizu, and Y.~Utsuno,
Prog.~Part.~Nucl.~Phys.~{\bf 47}, 319  (2001).  %pje Phys.
\bibitem{alex} B.~A.~Brown, 
Prog.~Part.~Nucl.~Phys.~{\bf 47}, 517  (2001),  %pje Phys.
and references therein.
\bibitem{oslo} M.~Hjorth-Jensen, T.~T.~S.~Kuo, and E.~Osnes,
Phys.~Rep.~{\bf 261}, 125 (1995).
\bibitem{matej1}M.~Lipoglav\v{s}ek {\it et al.}, Phys.~Rev.~C {\bf 65}, 
021302(R) (2002).
\bibitem{matej2}M.~Lipoglav\v{s}ek {\it et al.}, Phys.~Rev.~C {\bf 65}, 
051307(R) (2002).
\bibitem{matej3}M.~Lipoglav\v{s}ek {\it et al.}, Phys.~Rev.~C {\bf 66}, 
011302(R) (2002).
\bibitem{torgeir}T.~Engeland, M.~Hjorth-Jensen, and E.~Osnes,
    Phys.~Rev.~C {\bf 61}, 021302(R) (2000).

\bibitem{jo} J.~J.~Ressler, W.~B.~Walters, C.~N.~Davids, D.~J.~Dean, A.~Heinz, 
M.~Hjorth-Jensen, D.~Seweryniak, and J.~Shergur,
Phys.~Rev.~C {\bf 66}, 024308 (2002). 
\bibitem{anne} A.\ Holt, T. Engeland, M. Hjorth-Jensen, and E. Osnes,
    Nucl.~Phys.~{\bf A634}, 41 (1998).
\bibitem{bob1} S.~C.~Pieper, V.~R.~Pandharipande, R.~B.~Wiringa, and 
J.~Carlson, Phys.~Rev.~C {\bf 64}, 014001 (2001). 
\bibitem{bob2} S.~C.~Pieper, K.~Varga, and R.~B.~Wiringa, Phys.~Rev.~C 
{\bf 66}, 0044310 (2002). 
\bibitem{bob3}  R.~B.~Wiringa and S. C.~Pieper, Phys.~Rev.~Lett.~{\bf 89},
182501 (2002).
\bibitem{petr_erich2002} P.~Navr\'atil and W.~E.~Ormand, 
Phys. Rev. Lett.~{\bf 88}, 152502 (2002). 
\bibitem{petr_erich2003} P.~Navr\'atil and W.~E.~Ormand, 
Phys.~Rev.~C {\bf 68}, 034305 (2003). 
\bibitem{ko} T.~T.~S. Kuo and E.~Osnes, {\it Folded-diagram Theory 
of the Effective Interaction in  Atomic Nuclei}, Springer Lecture Notes 
in Physics (Springer, Berlin, 1990) Vol. 364;
T.~T.~S.\ Kuo, Lecture notes in
Physics; Topics in Nuclear Physics, eds.\ T.~T.~S.\ Kuo and S.~S.~M.\
Wong, (Springer, Berlin, 1981) Vol.\ {\bf 144}, p.\ 248.
\bibitem{igal} I.~Talmi, {\em Simple Models of Complex Nuclei},
Contemporary Concepts in Physics (Harwood Academic Publishers, 
Chur, Switzerland, 1993) Vol. 7.   %pje somewhat expanded
\bibitem{cc1} K.~Kowalski, D.~J.~Dean, M.~Hjorth-Jensen, T.~Papenbrock,
and P.~Piecuch, Phys.~Rev.~Lett.~{\bf 92}, 132501 (2004).
\bibitem{LMG65} H. J. Lipkin, N. Meshkov, and A. J. Glick, 
Nucl. Phys. {\bf 62}, 188 (1965).
\bibitem{vary04} K.~J.~Abraham and J.~P.~Vary, preprint nucl-th/0410037.
\bibitem{bran} B.~H.~Brandow, Rev.~Mod.~Phys.~{\bf39}, 771 (1967).
\bibitem{usrev} P.~J.~Ellis and E.~Osnes, Rev.~Mod.~Phys.~{\bf49}, 777 (1977).
\bibitem{suz} K.~Suzuki, R.~Okamoto, P.~J.~Ellis, and T.~T.~S.~Kuo,
Nucl.~Phys.~{\bf A567}, 576 (1994).
\end{thebibliography}
\end{document}